\documentclass[aps,prb,twocolumn,superscriptaddress]{revtex4-1}
\usepackage{amsmath,amsthm,amssymb}
\usepackage[dvipdfmx]{graphicx}
\usepackage{amsmath,bm}
\usepackage{color,ulem}
\usepackage{ifthen}
\usepackage{mdwlist}
\newcount \commentout
\newcommand{\1}{\mbox{1}\hspace{-0.25em}\mbox{l}}

\newlength{\figwidth}
\newlength{\figlarge}
\begin{document}
\title{
Reduction of topological $\mathbb{Z}$ classification in cold atomic systems
}
\author{Tsuneya Yoshida}
\affiliation{Department of Physics, Kyoto University, Kyoto 606-8502, Japan}
\author{Ippei Danshita}
\affiliation{Yukawa Institute for Theoretical Physics, Kyoto University, Kyoto 606-8502, Japan}
\author{Robert Peters}
\affiliation{Department of Physics, Kyoto University, Kyoto 606-8502, Japan}
\author{Norio Kawakami}
\affiliation{Department of Physics, Kyoto University, Kyoto 606-8502, Japan}
\date{\today}
\begin{abstract}
One of the most challenging problems in correlated topological systems is a realization of the reduction of topological classification, but very few experimental platforms have been proposed so far. 
We here demonstrate that ultracold dipolar fermions (e.g., $^{167}$Er, $^{161}$Dy, and $^{53}$Cr) loaded in an optical lattice of two-leg ladder geometry can be the first promising testbed for the reduction $\mathbb{Z}\to\mathbb{Z}_4$, where solid evidence for the reduction is available thanks to their high controllability. 
We further give a detailed account of how to experimentally access this phenomenon; around the edges, the destruction of one-particle gapless excitations can be observed by the local radio frequency spectroscopy, while that of gapless spin excitations can be observed by a time-dependent spin expectation value of a superposed state of the ground state and the first excited state. 
We clarify that even when the reduction occurs, a gapless edge mode is recovered around a dislocation, which can be another piece of evidence for the reduction.
\end{abstract}
\pacs{
***
} 
\maketitle

\textit{
Introduction.-
}
After the discovery of topological insulators, a topological perspective on condensed matter physics has become increasingly important~\cite{TI_review_Hasan10,TI_review_Qi10}.
The notion of topological phases has been extended to topological semi-metals and topological superconductors. Remarkably, these phases host exotic particles as low energy excitations, such as Weyl fermions, Majorana fermions \textit{etc.}, some of which have potential applications to quantum computations~\cite{Kitaev_chain_01,Alicia_Majorana_review12}.

The discovery of topological insulators has further brought great impact beyond solid state physics. In particular, it has provided a new arena of study in cold atoms, which is rapidly developing in these years~\cite{Zak_phase_cold_atoms_Atala_NatPhys13, Haldane_Jotzu_12, Topo_pump_Nakajima_16,Topo_pump_Lohse_16}. 
A significant advantage of cold atoms over materials is the high controllability, which has allowed unique observations for non-interacting topological systems, such as the Zak phase~\cite{Zak_phase_cold_atoms_Atala_NatPhys13}, the Thouless pump~\cite{Topo_pump_Nakajima_16,Topo_pump_Lohse_16}, and a symmetry-protected topological state~\cite{freeSPT_Song_arXiv17}.
With this remarkable success, it is not hard to imagine that the high controllability would be a key for solving one of the most significant issues in topological condensed matter physics, i.e. correlation effects on topological insulators/superconductors. 
Therefore, combining topology and strong correlations in cold atoms would provide a new perspective on correlated topological systems.

One of the striking phenomena induced by correlations in topological systems is  the reduction of topological classification.
Namely, correlation effects reduce the number of possible topological phases under certain symmetry classes. 
For instance, topological superconductors of symmetry class BDI follow $\mathbb{Z}$ classification in the absence of correlations while the systems follow $\mathbb{Z}_8$ classification in the presence of correlations~\cite{Z_to_Zn_Fidkowski_10}. 
In other words, eight Majorana fermions arising from the winding number $\nu=8$ are completely gapped out without symmetry breaking or gap-closing in the bulk.
Extensive studies on this issue~\cite{Turner11,YaoRyu_Z_to_Z8_2013,Ryu_Z_to_Z8_2013,Qi_Z_to_Z8_2013,Neupert_CS_2011,Lu_CS_2011,Levin_CS_2012,Fidkowski_Z162013,gu_supercohomology,Hsieh_CS_CPT_2014,Wang_Potter_Senthil2014,Metlitski_3Dinteraction2014,Wang_Senthil2014,You_Cenke2014,kapustin_fermionic_cobordisms2014,Isobe_Fu2015,Yoshida2015,Morimoto_2015, Superlattice_Yoshida17} have revealed that the reduction occurs in any dimension and is ubiquitous.

In spite of the above remarkable discovery, the following crucial question remains unsolved: \textit{How one can realize a testbed to observe the reduction of topological classification?}
The experimental observation is indispensable for further developments in correlated topological systems, and therefore such a feasible platform to observe the reduction is highly desired.
For solid evidence of the reduction, tuning the interaction is considered to be a key technique, but is rather difficult to control in the platform for solids~\cite{Superlattice_Yoshida17}. 
If one could find how to prepare such a platform, it would bring significant progress toward the observation of the reduction.
Unfortunately, however, very few experimental platforms have been proposed so far.

With this background, we tackle the above problem by focusing on cold atoms, in which system's parameters can be widely controlled. As a first step toward detection of the reduction, we here consider the simplest case, a one-dimensional correlated system.
Specifically, we demonstrate that ultracold dipolar fermions~\cite{Dy_dipolar_Lu,Er_dipolar_Aikawa14,Cr_dipolar_Reigue15}, e.g., $^{167}$Er, $^{161}$Dy, and $^{53}$Cr, loaded in a two-leg ladder optical lattice
serve as the first promising testbed of the reduction in one dimension, $\mathbb{Z}\to\mathbb{Z}_4$. 
Furthermore, we present a detailed account of how to experimentally access this phenomenon. 
In addition, we elucidate that gapless edge modes emerge around dislocations even when the reduction occurs, which can be another signal of the reduction.

\textit{
Reduction of topological classification in one-dimensional insulators, $\mathbb{Z}\to\mathbb{Z}_4$.-  
}
By employing a simple toy model, we first give an intuitive picture of the reduction in one dimension $\mathbb{Z}\to\mathbb{Z}_4$ in the presence of chiral symmetry
(for the definition, see Appendix~\ref{app: chiral symm}).
The corresponding symmetry class is AIII according to the Altland-Zirnbauer symmetry classes~\cite{Ryu_classification_free_2010}.
In the absence of correlations, one-dimensional topological insulators follow $\mathbb{Z}$ classification and are characterized by the winding number.
Recent studies based on the entanglement of the ground state~\cite{Turner11} or field theories~\cite{You_Cenke2014,Morimoto_2015} revealed that the classification result is reduced from $\mathbb{Z}$ to $\mathbb{Z}_4$ due to correlations.

\begin{figure}[!h]
\begin{center}
\includegraphics[width=\hsize,clip]{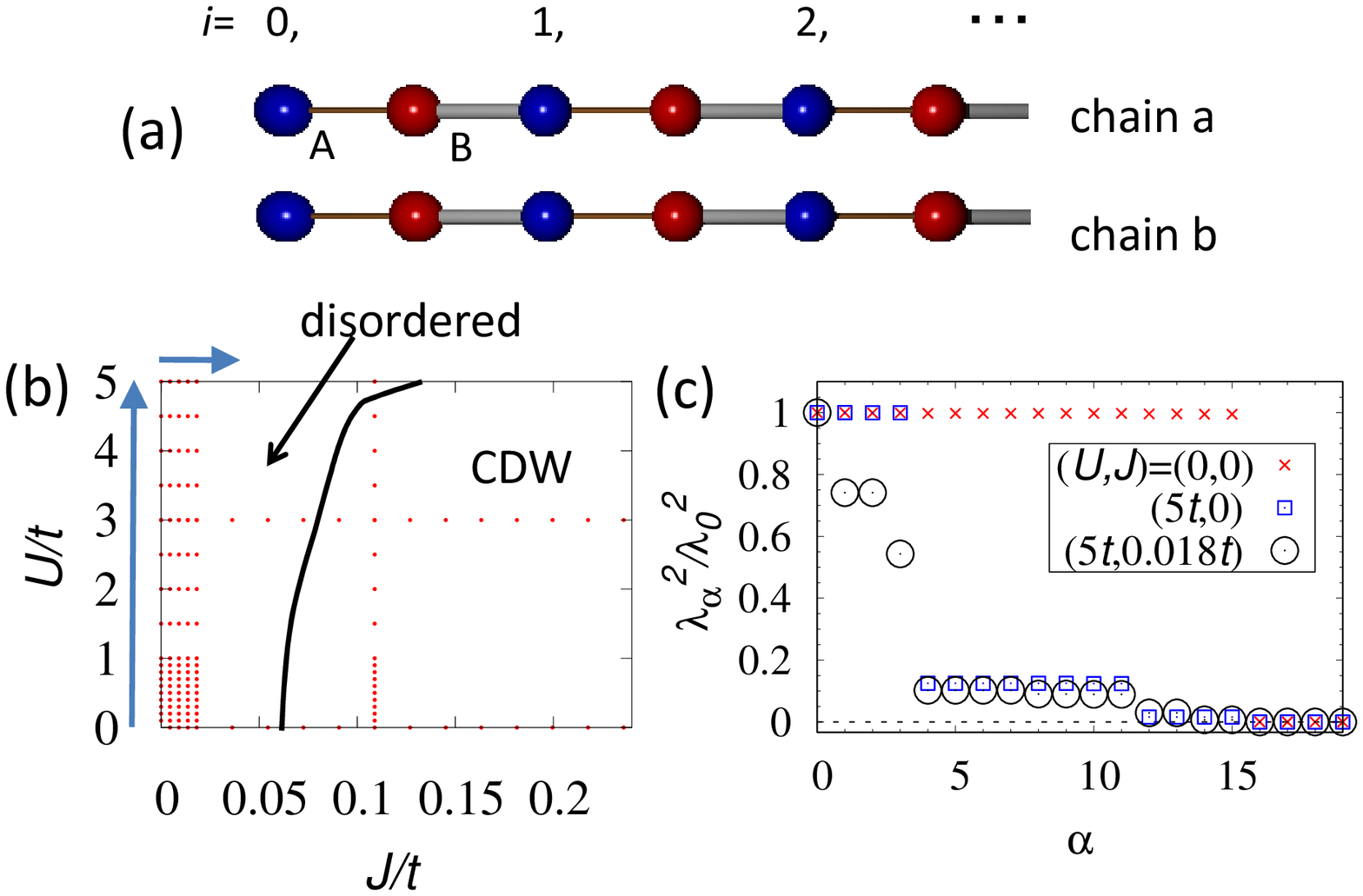}
\end{center}
\caption{(Color Online). 
(a): Sketch of the model~(\ref{eq: free_SSH}). Blue (red) circles denote $A$ ($B$) sublattice, respectively. 
Brown (gray) lines represent hopping $V$ ($t$), respectively.
(b): Phase diagram of the intra-ladder interaction $U$ vs. the inter-ladder interaction $J$. Red dots are data points.
(c): Highest 20 Schmidt eigenvalues for each case of parameters. Each eigenvalue is normalized so that the highest eigenvalue is one.
These data are obtained under the OBC and for $L=30$.
For calculation of the Schmidt eigenvalues ($\lambda_\alpha$), we consider a virtual cut dividing the system at the strong bond (gray line) with $i=L/2$.
}
\label{fig: OBC phase_ES}
\end{figure}

The reduction $\mathbb{Z}\to \mathbb{Z}_4$ can be observed by introducing interactions into the following two-leg Su-Schrieffer-Heeger (SSH) model composed of spin-half fermions,
\begin{eqnarray}
\label{eq: free_SSH}
H_0&=& -\sum_{i \alpha} (Vc^\dagger_{iA\alpha\sigma}c_{iB\alpha\sigma} +tc^\dagger_{i+1A\alpha\sigma}c_{iB\alpha\sigma}) +h.c.,
\end{eqnarray}
where $c^\dagger_{is\alpha\sigma}$ creates a fermion in spin-state $\sigma(=\uparrow,\downarrow)$ at sublattice $s(=A,B)$ and chain $\alpha(=a,b)$ of site $i$.
The lattice structure is shown in Fig.~\ref{fig: OBC phase_ES}(a). 
Gapless edge modes of the above model are expected to be unstable against interactions. The reason is as follows~\cite{Z4_DMFT_Yoshdia_17}. 
Introducing the intra-chain Hubbard interactions would destroy the gapless charge excitations and would leave spin excitations gapless. 
Further introducing appropriate inter-chain interactions, e.g., spin exchange interactions, would gap out the remaining gapless edge modes.
If the above argument indeed holds, it could verify the reduction $\mathbb{Z}\to\mathbb{Z}_4$~\cite{footnote_red}.

Now the problems to be solved are as follows.
(i) How one can implement the above model for the reduction $\mathbb{Z}\to\mathbb{Z}_4$ in cold atoms?
(ii) How one can observe the reduction in experiments?
Here, we naively think that introducing the kinetic spin exchange interaction of Heisenberg type between chains may be sufficient to realize the reduction. 
However, this scenario does not work because it breaks chiral symmetry~\cite{footnote_pertub} that is the key symmetry to be preserved in our study. 

\textit{
Dipolar fermions as a testbed of the reduction.-
}
In the following, we propose how to prepare a promising and feasible platform for observing the reduction experimentally.
Firstly, we note that the non-interacting part of the above model is considered to be feasibly prepared with optical lattices~\cite{footnote_SSH,Feolling_free_SSH07,Strabley_two-leg_06,Danshita_two_leg_07,Chen_two_leg_10}.

Now, we discuss how to prepare a system with chiral symmetry where fermions with (pseudo-)spin half interacts with each other by spin-exchange interactions.
We find that this is accomplished by employing dipolar fermions (e.g., $^{167}$Er, $^{161}$Dy, and $^{53}$Cr). 
Here, specifically, consider two ${}^{161}\mathrm{Dy}$ atoms, labeled by 1 and 2.
These atoms interact with each other via the magnetic dipole-dipole interaction~\cite{Dy_dipolar_Lu}
\begin{eqnarray}
\label{eq: dipole}
U_{dd}&=&\frac{\mu_0(2\mu_B)^2}{4\pi r^3} [\bm{S}_1\cdot \bm{S}_2 -\frac{3}{r^2}(\bm{S}_1\cdot\bm{r}) (\bm{S}_2\cdot\bm{r} )],
\end{eqnarray}
where $\bm{r}:=\bm{r}_1-\bm{r}_2$, and $\bm{r}_{1(2)}$ denotes the position vector of atom $1(2)$. $\bm{S}_{1(2)}$ denotes the total spin operator of electrons in the atom $1(2)$, respectively.
$\mu_0$ denotes the permeability of vacuum. $\mu_B$ denotes the Bohr magneton.
Thus, loading ${}^{161}\mathrm{Dy}$ atoms, one can prepare a system where fermions interact with each other via the magnetic dipole-dipole interactions.
However, just loading the dipolar fermions is not sufficient,  because ${}^{161}\mathrm{Dy}$ atoms have huge spin $F=21/2$, where $F$ denotes the total spin of nuclear and electronic spins. Therefore, one has to restrict the Hilbert space spanned by the states with $m_F=21/2,19/2,\cdots,-21/2$ to the subspace spanned by two states, e.g. $m_F=21/2, 19/2$, where $m_F$ denotes the $z$-component of the spin.

The restriction of the Hilbert space is accomplished by the following three steps.
(i) 
Prepare atoms in the states with $m_F=\pm21/2,\pm19/2$ by applying the optical pumping~\cite{Happer_opticalpump_RMP} which excites the states with $F=21/2$ to the states with $F=17/2$.
(ii) 
Remove atoms in the states with $m_F=-21/2,-19/2$ by temporarily applying a magnetic field.
(iii) 
Continue to shine the laser in the first step to forbid the transition via the dipolar relaxation~\cite{Dipolar_relax_Hensler2003} to the other states with $m_F=17/2,15/2,\cdots,-21/2$. The transition can be prevented due to the quantum Zeno effect~\cite{Schafer_Zenoeffect}.
We refer to the state with $m_F=21/2$ ($m_F=19/2$) 
as an effective up- (down-) spin state, respectively.
Note that the intra-chain Hubbard interaction can be tuned by Feshbach resonance~\cite{Feshbach_1958,Baumann_Dy_Feshbach14}. 

We thus end up with the following effective Hamiltonian:
\begin{subequations}
\label{eq: effective H}
\begin{eqnarray}
H&=&H_0+U\sum_{i\alpha}(n_{is\alpha\uparrow}-\frac{1}{2})(n_{is\alpha\downarrow}-\frac{1}{2})+J\sum_{i}h_{is}, \nonumber \\
\end{eqnarray}
\begin{eqnarray}
h_{is}&=&
A_1(\tilde{S}^x_{isa}\tilde{S}^x_{isb}+\tilde{S}^y_{isa}\tilde{S}^y_{isb})
-A_2\tilde{S}^z_{isa}\tilde{S}^z_{isb} \nonumber \\ 
&&
-A_3(n_{isa}-1)(n_{isb}-1) \nonumber \\ 
&&
-A_4[(n_{isa}-1)\tilde{S}^z_{isb} +(n_{isb}-1)\tilde{S}^z_{isa}],
\end{eqnarray}
\end{subequations}
where $A_1=16^2/21$, $A_2=16/21$, $A_3=2 \times(160/21)^2$, $A_4=(20\times 16^2)/(21^2)$, respectively. Here, $\tilde{\bm{S}}$'s are pseudo-spin operators acting on the Hilbert space with $F=21/2$ and $m_F=21/2,19/2$. 
We have assumed that the two chains are aligned along the $z$-direction, and that the distance between adjacent sites in the same chain is sufficiently large, allowing us to neglect the dipole-dipole interaction in the same chain. The detail of the derivation is given in Appendix~\ref{app: projection}. 
Note that this system respects the chiral symmetry (see Appendix~\ref{app: chiral symm}). 
In experiments, the strength of the dipole-dipole interaction can be tuned by changing the distance between chains, $l_0$. 
The maximum strength is estimated to be $U_{dd}\lesssim 0.1t$ with $t\sim 1.0\mathrm{kHz}$ and $l_0\sim266\mathrm{nm}$~\cite{Baier_Science15}. 
Thus, a realistic value of $J$ in experiments is approximately $0\lesssim J \lesssim 0.01t$.

\textit{
Density-matrix renormalization group (DMRG) simulations for reduction: bulk and edge properties.-
}
Now, using the DMRG method~\cite{White_DMRG_92,Schollwoeck_DMRG_05,Schollwoeck_DMRG_11}, we demonstrate that the reduction of topological classification occurs in our system.
Let us start with the phase diagram of the intra-chain Hubbard interaction $U$ vs. the inter-chain interaction $J$ [Fig.~\ref{fig: OBC phase_ES}(b)]. The phase diagram is obtained for $V=0.1t$ under the open boundary condition (OBC).
Unless otherwise noted, we set $V=0.1t$ in the following.
For small $J$, the system is in the disordered phase while for large $J$, the system is in the charge-density-wave (CDW) phase. 
The chiral symmetry is preserved in the disordered phase, while it is broken in the CDW phase.
We note that the CDW order breaks discrete symmetry and thus does not contradict the Mermin-Wagner theorem. The CDW order is induced by inter-chain density-density interaction in Eq.~(\ref{eq: effective H}b).

Here, we note that under the periodic boundary condition (PBC), the system is gapped for $0\leq U\leq5t$ with $J=0$ and for $0\leq J\leq 0.018t$ with $U=5t$, respectively. 
These parameter regions are indicated as blue arrows in the phase diagram [see Fig.~\ref{fig: OBC phase_ES}(b)].
Namely, the charge gap ($\Delta_c$) and the spin gap ($\Delta_s$) are open for these parameter sets, where the gaps are defined as $\Delta_c=E_{2L+1,1/2}- E_{2L,0}$ and $\Delta_s=E_{2L,1}- E_{2L,0}$ ($L$ denotes the length of the chain),  respectively. Here,
$E_{N,\tilde{S}^z}$ denotes the lowest energy of the Hilbert space labeled by the total number of fermions and $z$-component of the total pseudo-spin. 
We note that the system preserves the total number of fermions for each species. For more detail of the bulk properties, see Appendix~\ref{app: PBC}.

The reduction occurs in the disordered phase. 
Let us first observe the reduction via the degeneracy of the entanglement spectrum (ES) which is calculated in the bulk.
Via the ES in the bulk one can deduce topological properties of the system; the degeneracy of the lowest entanglement energy states predicts the emergence of gapless modes around the edges~\cite{Li_ES_PRL08}.
In the following, we observe that the degeneracy of the ES is lifted as the interactions $U$ and $J$ are turned on.
The 20 lowest Schmidt eigenvalues, ${\lambda^2}_\alpha$, are plotted in Fig.~\ref{fig: OBC phase_ES}(c) for several cases of parameters. 
The entanglement energy can be read off from the corresponding Schmidt eigenvalue via $E_\alpha=-2\log(\lambda_\alpha)$.
We note that this spectrum is obtained under the OBC and for a virtual cut.
In the non-interacting case [see data of $(U,J)=(0,0)$ in Fig.~\ref{fig: OBC phase_ES}(c)], the ES shows the 16-fold degeneracy in accordance with nontrivial topological properties of free fermions; the 16-fold degeneracy indicates gapless edge modes in the single-particle spectrum for each channel $(\alpha,\sigma)$, which is consistent with the winding number taking one for each channel $(\alpha,\sigma)$.
Turning on the repulsive Hubbard interaction $U$ lifts the degeneracy from 16-fold to 4-fold [see data for $(U,J)=(5t,0)$ in Fig.~\ref{fig: OBC phase_ES}(c)], indicating the emergence of gapless excitations only in a collective excitation spectrum. 
As we see below, these modes emerge in the spin excitation spectrum.
Furthermore, turning on the inter-chain coupling $J$ completely lifts the degeneracy of the ES; no degeneracy is observed for $(U,J)=(5t,0.018t)$ in Fig.~\ref{fig: OBC phase_ES}(c).
Correspondingly, fermions in chain $a$ and $b$ form a singlet at each site.
In the above, we have seen that introducing interactions, $U$ and $J$, lifts the 16-fold degeneracy of the ES without chiral symmetry breaking. 
This result indicates the reduction of topological classification  $\mathbb{Z}\to\mathbb{Z}_4$.

\begin{figure}[!h]
\begin{center}
\includegraphics[width=\hsize,clip]{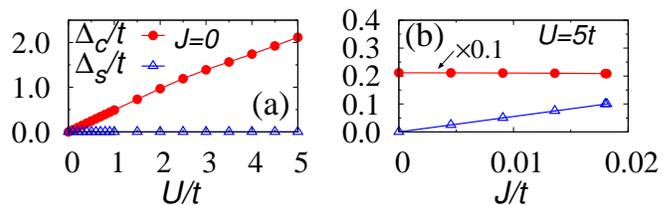}
\end{center}
\caption{(Color Online). 
(a) [(b)]: The charge gap $\Delta_c$ and the spin gap $\Delta_s$ as functions of the interaction strength under the OBC for $J=0$ ($U=5t$), respectively.
In panel (b), the data of $\Delta_c$ is multiplied by $0.1$.
}
\label{fig: OBC gap}
\end{figure}

Now let us turn to the edge properties.
In the following, we demonstrate that gapless edge modes of the non-interacting case are completely gapped out without chiral symmetry breaking. 
Since we have confirmed the presence of the gap in the particle excitation spectrum and the spin excitation spectrum under the PBC, the origin of gapless excitation under the OBC indicates the presence of gapless edge modes.
For $(U,J)=(0,0)$, we can see that both of the charge and spin gaps are zero, indicating the gapless excitation of the single-particle spectrum [Fig.~\ref{fig: OBC gap}(a)]. 
Switching on the interaction $U$ opens the charge gap and keeps the spin gap zero, indicating the emergence of gapless edge modes only in the spin excitation spectrum [Fig.~\ref{fig: OBC gap}(a)]. 
Furthermore the introduction of $J$ destroys the remaining gapless edge modes in the spin excitation spectrum [Fig.~\ref{fig: OBC gap}(b)]. 
The above data under the OBC indicate that all of the gapless edge modes are destroyed in this model.

With all the above numerical results of the ES and the gaps under the OBC, we come to the conclusion that our model~(\ref{eq: effective H}) simulates the reduction of topological classification, $\mathbb{Z}\to\mathbb{Z}_4$.

\textit{
How to observe the reduction in experiments.-
}
Our numerical simulation indicates that the opening of the charge and spin gaps at edges is a signal of the reduction.
Now, the remaining problem we have to address is how to observe these excitation gaps.

To observe a gap opening of edge modes in the single-particle spectrum, we can make use of the local radio-frequency spectroscopy~\cite{Fukuhara_local_radio_13}.
The gap size is estimated to be $\Delta_c\sim2\mathrm{kHz}$. 

On the other hand, to detect the gap formation in spin excitations, a more elaborated method is necessary. We find that it can be extracted from time-evolution of a superposed state composed of the ground state and the first excited state~\cite{ST-osci_Grief13}.
The basic idea is as follows~\cite{footnote_time_evl}. 
%
Consider a wave function $|\psi(0)\rangle$ composed of a linear combination of the ground state $|1\rangle$ and an excited state $|2\rangle$,
$|\psi(0)\rangle= c_1 |1\rangle + c_2 |2\rangle$ with $c_1$, $c_2\in \mathbb{C}$.
Under the time evolution, the state is written as
$|\psi(t)\rangle = c_1 e^{-iE_1t}   |1\rangle + c_2 e^{-iE_2t} |2\rangle$.
Thus, the expectation value of an operator $A$ is written as 
\begin{eqnarray}
\label{eq: At}
\langle A(t) \rangle &=& \sum_i |c_i|^2 \langle i | A | i \rangle +2a_{12}\cos[\omega_{21}t+\delta_{12}],
\end{eqnarray}
with $a_{12}e^{i\delta_{12}}:= c^*_1 c_2 \langle 1| A |2\rangle$, $a_{12}>0$.
By measuring the frequency, one can read off the size of the gap $\omega_{21}:=E_2-E_1$. 
Based on this prescription, one can observe the spin gap by (i) shining a half-$\pi$ pulse only to the chain $a$ and (ii) observing frequency of $\langle \tilde{S}^x_a(t)\rangle$ under the time-evolution with the Hamiltonian~(\ref{eq: effective H}). 
Here, we explain the details of each step. 
First, shining the half-$\pi$ pulse maps the singlet to the superposed state
\begin{eqnarray}
\frac{1}{\sqrt{2}}|singlet\rangle
+
\frac{i}{2}[ |\downarrow\rangle_a|\downarrow\rangle_b-|\uparrow\rangle_a|\uparrow\rangle_b],
\end{eqnarray}
where $|\sigma\rangle_{a(b)}$ with $\sigma=\uparrow,\downarrow$ describes the spins around the edges of chain $a$ ($b$), respectively. 
$|singlet\rangle:=(|\uparrow\rangle_a|\downarrow\rangle_b-|\downarrow\rangle_a|\uparrow\rangle_b)/\sqrt{2}$. 
The  numerical results show that the energy of the triplet state with $\tilde{S}^z=1$ and that with $\tilde{S}^z=-1$ is identical (see Appendix~\ref{app: Emin_up_down}). Second, in real experiments, $\langle \tilde{S}^x_a(t)\rangle$ can be measured by applying a half-$\pi$ pulse (along $\tilde{S}^y$-axis) for both channels;
\begin{eqnarray}
\langle \psi(t)|  \Pi^\dagger_{1/2}\tilde{S}^z_a \Pi_{1/2} | \psi(t)\rangle
&&=\langle \psi(t)| \tilde{S}^x_a | \psi(t)\rangle,
\end{eqnarray}
with $\Pi_{1/2}=\mathrm{exp}[i\pi(\sigma^x_a+\sigma^x_b)/4]$ arising from the half-$\pi$ pulse. The matrices $\sigma_{a(b)}$'s denote the Pauli matrices acting on a fermion in pseudo-spin state of chain $a$ ($b$), respectively.

The numerical simulation shows that the gap size is approximately $\Delta_s \simeq 0.1 t \sim 100\mathrm{Hz}$. 
Hence, at the lowest temperature achieved in two-component fermion systems, which is $T\simeq 0.25t$~\cite{Mazurenko_AForder_DMD2017}, thermal fluctuations significantly mix the singlet ground state with the excited states at edges.
We can, however, prepare the singlet edge state by making use of feedback control~\cite{Feedbackcontrol_Inoue_PRL13,Nondemolition_Yamamoto_17} and a singlet-triplet oscillation~\cite{ST-osci_Grief13} (see Appendix~\ref{app: purifying}).
In this way, by direct observation of excitation gaps under the OBC, we can access the reduction $\mathbb{Z}\to\mathbb{Z}_4$ in cold atoms.

\begin{figure}[!h]
\vspace{5mm}
\begin{center}
\includegraphics[width=60mm,clip]{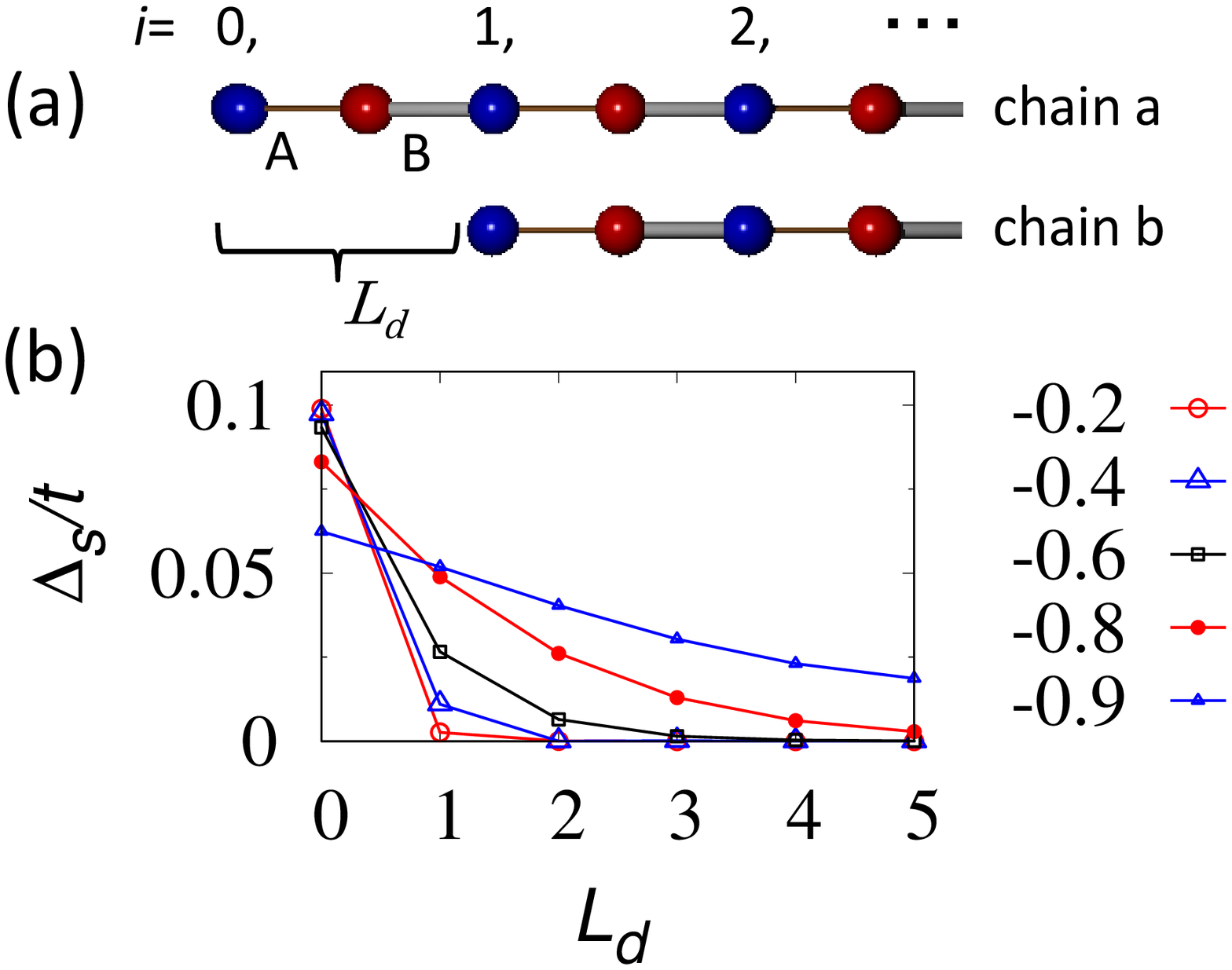}
\end{center}
\caption{(Color Online). 
(a): Sketch of a dislocation with $L_d=1$.
(b): The spin gap $\Delta_s$ for $V=-0.2t$, $-0.4t$, $-0.6t$, $-0.8t$, and $-0.9t$ in the presence of dislocations.
The data is obtained for $(U,J)=(5t,0.018t)$.
}
\label{fig: disloc}
\end{figure}

\textit{
Effects of dislocations on the reduction.-
}
We now demonstrate that even when the reduction occurs, a gapless edge mode emerges around a dislocation, which can be another clear signal of the reduction.
Let us consider a system with a dislocation characterized with $L_d$ [Fig.~\ref{fig: disloc}(a)].
In this case, the chain $a$ does not couple with the chain $b$ for $0<i_x<L_d$, meaning that a single chain with the Hubbard interaction emerges for $i<L_d$. We recall that the inter-chain coupling is essential for gapping out edge modes in the spin excitation spectrum.
Thus, a gapless edge mode emerges only in the spin excitation spectrum, which is reminiscent of edge states in the so-called topological Mott insulator~\cite{TMI_LBalents09,TMI_1D_yoshida,TMI_2D_yoshida,Zhou_TMI_Yb_PRL17}.
The results shown in Fig.~\ref{fig: disloc}(b) support this scenario. The spin gap is plotted for several values of $V$. In this figure, we can see that the gap size decreases and finally becomes zero with increasing $L_d$, indicating the emergence of the gapless edge mode.

The above results elucidate that the presence of decoupled gapless edge modes only around the dislocations can be other definite evidence for the reduction. We believe that this argument can be extended to higher dimensions.

\textit{
Conclusion.-
}
We have proposed a promising experimental testbed for a realization of the reduction of topological classifications, which is one of the most challenging problems in correlated topological systems. The proposed setup with cold atoms allows us to turn on/off interactions in experiments, making distinct evidence available.
The experimental platform can be implemented by loading ultracold dipolar fermions, e.g., ${}^{161}\mathrm{Dy}$ atoms, into the two-leg SSH model and by making use of the quantum Zeno effect. We have also demonstrated how to observe the reduction experimentally, which can be feasibly done by direct measurements of energy gaps with the Radio frequency spectroscopy and the time evolution of superposed states. Furthermore, we have pointed out that when the reduction occurs, dislocations host gapless edge modes only in the spin excitation spectrum, which are reminiscent of a topological Mott insulator. The emergence of such gapless edge modes only around dislocation can be another piece of solid evidence of the reduction.

The results in this paper are expected to serve as a foothold for the experimental observation of the reduction in higher dimensions, and also for other exotic correlated topological systems, such as interaction enabled topological crystalline phases~\cite{Lapa_16}.

\textit{acknowledgements.-}
This work is partly supported by JSPS KAKENHI Grant No. 25220711, JP15H05855, and No. JP16K05501, and CREST, JST No. JPMJCR1673. The numerical calculations were performed on supercomputer at the ISSP in the University of Tokyo, and the SR16000 at YITP in Kyoto University.

%

\clearpage

\renewcommand{\thesection}{S\arabic{section}}
\renewcommand{\theequation}{S\arabic{equation}}
\setcounter{equation}{0}
\renewcommand{\thefigure}{S\arabic{figure}}
\setcounter{figure}{0}
\renewcommand{\thetable}{S\arabic{table}}
\setcounter{table}{0}
\makeatletter
\c@secnumdepth = 2
\makeatother

\onecolumngrid
\begin{center}
 {\large \textmd{Supplemental Materials:} \\[0.3em]
 {\bfseries Reduction of topological $\mathbb{Z}$ classification in cold atomic systems}}
\end{center}

\setcounter{page}{1}

\section{
derivation of effective interactions
}
\subsection{
Projecting the dipolar interaction
}\label{app: projection}

Here, we show that the magnetic dipole-dipole interaction yields Eq.~(\ref{eq: effective H}b).
Consider two ${}^{161}\mathrm{Dy}$ atoms aligned along the $z$-direction. (We label these two atoms as 1 and 2.)
The magnetic dipole-dipole interaction between electrons in the atoms is given by
\begin{eqnarray}
U_{dd}&=&\frac{\mu_0(2\mu_B)^2}{4\pi r^3} [\bm{S}_1\cdot \bm{S}_2 -3S^z_1S^z_2],
\end{eqnarray}
with $\bm{r}:=\bm{r}_1-\bm{r}_2$. $\bm{S}_1$ and $\bm{S}_2$ denote the total spin operators of electrons in the dipolar atoms. 
$\mu_0$ denotes the permeability of vacuum. $\mu_B$ denotes the Bohr magneton.

Here, we calculate the effective interaction terms for the subspace spanned by the following two states
\begin{subequations}
\begin{eqnarray}
|F=21/2,m_F=21/2\rangle &=& |I=5/2,I^z=5/2\rangle\otimes|S=8,S^z=8\rangle := |I^z=5/2,S^z=8\rangle,
\end{eqnarray}
and 
\begin{eqnarray}
|F=21/2,m_F=19/2\rangle &=& \sqrt{\frac{5}{21}}|I=5/2,I^z=3/2\rangle\otimes|S=8,S^z=8\rangle + \frac{4}{\sqrt{21}}|I=5/2,I^z=5/2\rangle\otimes|S=8,S^z=7\rangle, \nonumber \\
&:=&
\sqrt{\frac{5}{21}}|I^z=3/2,S^z=8\rangle + \frac{4}{\sqrt{21}}|I^z=5/2,S^z=7\rangle,
\end{eqnarray}
\end{subequations}
where $|I,I^z\rangle$ denotes the state with total nuclear spin $I$ and $z$-component of the spin $I^z$. $|S,S^z\rangle$ denotes the state with total electronic spin $S$ and the $z$-component $S^z$.
For simplicity we rewrite the states $(|F=21/2,m_F=21/2\rangle,|F=21/2,m_F=19/2\rangle)$ as $(|m_F=21/2\rangle,|m_F=19/2\rangle)$ unless otherwise noted.

The projection operator to this subspace is given by
\begin{eqnarray}
\label{eq: proj}
P&=&|m_F=21/2\rangle \langle m_F=21/2| \nonumber \\
 &&+ |m_F=19/2\rangle \langle m_F=19/2|.
\end{eqnarray}

Here we derive effective interactions in this subspace.
The building block of the effective interactions is $PS^sP$'s with $s=x,y,z$ because  
\begin{eqnarray}
\bm{S}_1\cdot \bm{S}_2 &\to& P_1P_2\bm{S}_1\cdot \bm{S}_2P_1P_2= P_1\bm{S}_1P_1\cdot P_2\bm{S}_2P_2,
\end{eqnarray}
holds,
where $P_1$ ($P_2$) is the projection operator [Eq.~(\ref{eq: proj})] acting on the atom 1 (2).

With $c=\sqrt{\frac{5}{21}}$ and $c'=\frac{4}{\sqrt{21}}$, applying the operator $P$ to the spin operator $S^z$ yields
\begin{subequations}
\begin{eqnarray}
&&P S^z P \nonumber \\
&&=P[  8|m_F=21/2\rangle \langle m_F=21/2| \nonumber \\
&& \quad +(8c|I_z=3/2,S_z=8\rangle \nonumber \\
&& \quad +7c'|I_z=5/2,S_z=7\rangle)\langle m_F=19/2|], \nonumber \\
&&=8|m_F=21/2\rangle \langle m_F=21/2| \nonumber \\
&& \quad +(8|c|^2+7|c'|) |m_F=19/2 \rangle \langle m_F=19/2|, \nonumber \\
&&=\frac{8+8|c|^2+7|c'|}{2} P \nonumber \\
&& \quad +\frac{8-(8|c|^2+7|c'|)}{2}(| m_F=21/2\rangle \langle m_F=21/2| \nonumber \\
&& \quad \quad \quad \quad  \quad \quad  -|m_F=19/2\rangle \langle m_F=19/2|). 
\end{eqnarray}
With the second quantization, the above term is written as
\begin{eqnarray}
&&P S^z P \nonumber \\
&&\to A (n_\uparrow+n_\downarrow)+B\tilde{S}^z,
\end{eqnarray}
\end{subequations}
where we have defined pseudo-spin as $(| m_F=21/2\rangle, | m_F=19/2\rangle):=(|\uparrow \rangle, |\downarrow \rangle)$.  
$n_\sigma$ ($\sigma=\uparrow,\downarrow$) denotes the corresponding density operator of the pesudo-spin state. $\tilde{S}$'s are pseudo-spin operators.
We have defined $A$ and $B$ as $A=(8+8|c|^2 +7|c'|^2)/2=160/21$ and $B=8-(8|c|^2 +7|c'|^2)=16/21$, respectively.

\begin{eqnarray}
&&P S^+ P \nonumber \\
&&=P  4c'|I_z=5/2,S_z=8\rangle \langle F=21/2,m_F=19/2|, \nonumber \\
&&=4c'|F=21/2,m_F=21/2\rangle \langle F=21/2,m_F=19/2|, \nonumber \\
&&=C\tilde{S}^+,
\end{eqnarray}
with $C=4c'=\frac{16}{\sqrt{21}}$.

\begin{eqnarray}
&&P S^- P \nonumber \\
&&=P  4|I_z=5/2,S_z=7\rangle \langle F=21/2,m_F=21/2|, \nonumber \\
&&=4c'^*  |F=21/2,m_F=19/2\rangle \langle F=21/2,m_F=21/2|, \nonumber \\
&&=C^*\tilde{S}^-.
\end{eqnarray}

Thus, under projection, each term is written as
\begin{eqnarray}
&& S^z_iS^z_j \nonumber \\
&&\to (A\sum_\sigma n_{i\sigma} +B\tilde{S}^z_i) (A\sum_\sigma n_{j\sigma} +B\tilde{S}^z_j), \nonumber \\
&&= A^2\sum_\sigma n_{i\sigma}\sum_{\sigma'} n_{i\sigma'}+AB\tilde{S}^z_i \sum_\sigma n_{j\sigma} \nonumber \\
&& \quad \quad \quad + AB \sum_\sigma n_{i\sigma}\tilde{S}^z_j +B^2\tilde{S}^z_i\tilde{S}^z_j,
\end{eqnarray}

\begin{eqnarray}
&& S^x_iS^x_j \nonumber \\
&&=\frac{1}{2^2} (S^+_i+S^-_i)(S^+_j+S^-_j), \nonumber \\
&&= \frac{1}{2^2} (S^+_i S^+_j +S^-_iS^-_j) +\frac{1}{2^2} (S^+_iS^-_j +S^-_i S^+_j), \nonumber \\
&&\to \frac{1}{2^2}(C^2S^+_i S^+_j +C^*{}^2 S^-_iS^-_j)
     +\frac{|C|^2}{2^2}( S^+_iS^-_j +  S^-_i S^+_j), \nonumber \\
\end{eqnarray}

\begin{eqnarray}
&& S^y_iS^y_j \nonumber \\
&&= -\frac{1}{2^2} (S^+_i-S^-_i)(S^+_j-S^-_j), \nonumber \\
&&= \frac{-1}{2^2} (S^+_i S^+_j +S^-_iS^-_j) +\frac{1}{2^2} (S^+_iS^-_j +S^-_i S^+_j), \nonumber \\
&&\to \frac{-1}{2^2}(C^2S^+_i S^+_j +C^*{}^2 S^-_iS^-_j)
     +\frac{|C|^2}{2^2}( S^+_iS^-_j +  S^-_i S^+_j). \nonumber \\
\end{eqnarray}

Therefore, we find that in the subspace, the magnetic dipole-dipole interaction is projected as 
\begin{eqnarray}
\label{eq: projected U}
U_{dd}&=&
\frac{\mu_0(2\mu_B)^2}{4\pi r^3}
[
S^x_iS^x_j+S^y_iS^y_i -2 S^z_iS^z_i
],
\nonumber \\
&&\to
\frac{\mu_0(2\mu_B)^2}{4\pi r^3}
[
C^2(\tilde{S}^x_i\tilde{S}^x_j+\tilde{S}^y_i\tilde{S}^y_i)
-2B^2\tilde{S}^z_i\tilde{S}^z_j
\nonumber \\
&&\quad\quad
-2A^2n_{i}n_{j}-2AB\tilde{S}^z_i n_{j} -2AB n_{i}\tilde{S}^z_j
],
\end{eqnarray}
with $n_{i}:=\sum_{\sigma}n_{i\sigma}$.

The cold atoms considered in the main text conserve the total number of particles and the $z$-component of total pseudo-spin. Thus, we can add the terms, $\sum_{i} n_{i}$ and $\sum_i \tilde{S}^z_i$ with changing the origin of the total energy.
Therefore, the effective interaction is written as
\begin{eqnarray}
U_\mathrm{eff}&=&
\frac{\mu_0(2\mu_B)^2}{4\pi r^3}
[
C^2(\tilde{S}^x_i\tilde{S}^x_j+\tilde{S}^y_i\tilde{S}^y_i)
-2B^2\tilde{S}^z_i\tilde{S}^z_j
\nonumber \\
&&\quad\quad\quad\quad
-2A^2(n_{i}-1)(n_{j}-1)-2AB\tilde{S}^z_i (n_{j}-1) \nonumber \\
&&\quad\quad\quad\quad
 -2 AB (n_{i}-1)\tilde{S}^z_j
].
\end{eqnarray}
Here we have added the following term which is reduced to a constant value:
\begin{eqnarray}
-2A^2(n_{i}+n_{j}+1) +2AB\tilde{S}^z_i +2AB\tilde{S}^z_j.
\end{eqnarray}
We note that the dipolar interaction between atoms in the same chain is negligible in the limit of large distance between sites.

\subsection{
Chiral symmetry breaking term and how to restore the symmetry
}\label{app: chiral symm}

Here, we show that the system preserves the chiral symmetry.
The chiral transformation is written as 
\begin{subequations}
\begin{eqnarray}
\Xi&=& U_\Xi \mathcal{K},
\end{eqnarray}
with 
\begin{eqnarray}
U_\Xi&=& \Pi_{is\alpha} (c_{is\alpha\uparrow}+ \mathrm{sgn}(s) c^\dagger_{is\alpha\uparrow})(c_{is\alpha\downarrow}+ \mathrm{sgn}(s) c^\dagger_{is\alpha\downarrow}). \nonumber \\
\end{eqnarray}
\end{subequations}
$\mathrm{sgn}(s)$ takes 1 (-1) for $s=A$ $(s=B)$, respectively.
We note that $\Xi$ satisfies $\Xi^2=1$.

Thus, annihilation operators are transformed as follows:
\begin{eqnarray}
\Xi c_{is\alpha\sigma} \Xi^{-1}
&=&
\mathrm{sgn}(s)c^\dagger_{is\alpha\sigma}.
\end{eqnarray}

In the following, we clarify how each term of Eq.~(\ref{eq: projected U}) is transformed under applying the operator $\Xi$.

Concerning the spin exchange interactions
\begin{subequations}
\begin{eqnarray}
\Xi \tilde{S}^s_{iAa} \tilde{S}^s_{iAb} \Xi^{-1}&=& \Xi \tilde{S}^s_{iAa} \Xi \Xi^{-1}\tilde{S}^s_{iAb} \Xi^{-1}, \nonumber\\
&=& \tilde{S}^s_{iAa} \tilde{S}^s_{iAb},
\end{eqnarray}
holds for $s=x,y,z$
because 
\begin{eqnarray}
\Xi \tilde{S}^s_{iA\alpha}  \Xi^{-1}&=& \Xi \frac{1}{2}c^\dagger_{iA\alpha\sigma}\sigma^s_{\sigma\sigma'} c_{iA\alpha\sigma'} \Xi^{-1}, \nonumber \\
&=& \frac{1}{2} c_{iA\alpha\sigma}\sigma^s_{\sigma\sigma'} c^\dagger_{iA\alpha\sigma'}, \nonumber \\
&=& \frac{1}{2} c^\dagger_{iA\alpha\sigma'} \sigma^s_{\sigma'\sigma}{}^T c_{iA\alpha\sigma}  +\mathrm{tr}\sigma^s, \nonumber \\
&=& \frac{1}{2} c^\dagger_{iA\alpha\sigma'} \sigma^s_{\sigma'\sigma}{}^T c_{iA\alpha\sigma},
\end{eqnarray}
\end{subequations}
holds. 
$\sigma$'s are the Pauli matrices acting on the pseudo-spin space.
We note that the same calculation holds for sublattice B.

The other terms are transformed as follows.
\begin{eqnarray}
&&\Xi [(n_{isa\uparrow}+n_{isa\downarrow}-1)\tilde{S}^z_{isb}+ (a\leftrightarrow b)]\Xi^{-1} \nonumber \\
&=&(1-n_{isa\uparrow}-n_{isa\downarrow})(-\tilde{S}^z_{isb})+ (a\leftrightarrow b), \nonumber \\
&=& [(n_{isa\uparrow}+n_{isa\downarrow}-1)\tilde{S}^z_{isb}+ (a\leftrightarrow b)],
\end{eqnarray}
and 
\begin{eqnarray}
&&\Xi [(n_{isa\uparrow}+n_{isa\downarrow}-1) (n_{isb\uparrow}+n_{isb\downarrow}-1) ]\Xi^{-1} \nonumber \\
&=& [1-(n_{isa\uparrow}+n_{isa\downarrow})][1-(n_{isb\uparrow}+n_{isb\downarrow}) ],  \nonumber \\
&=& [n_{isa\uparrow}+n_{isa\downarrow} -1 ][n_{isb\uparrow}+n_{isb\downarrow}-1 ],
\end{eqnarray}
for $s=A,B$.

Therefore, all the terms in the model~(\ref{eq: effective H}) is invariant under the chiral transformation.

\section{
Properties under the periodic boundary conditions
}
\label{app: PBC}
\begin{figure}[!h]
\begin{minipage}{0.47\hsize}
\begin{center}
\includegraphics[width=50mm,clip]{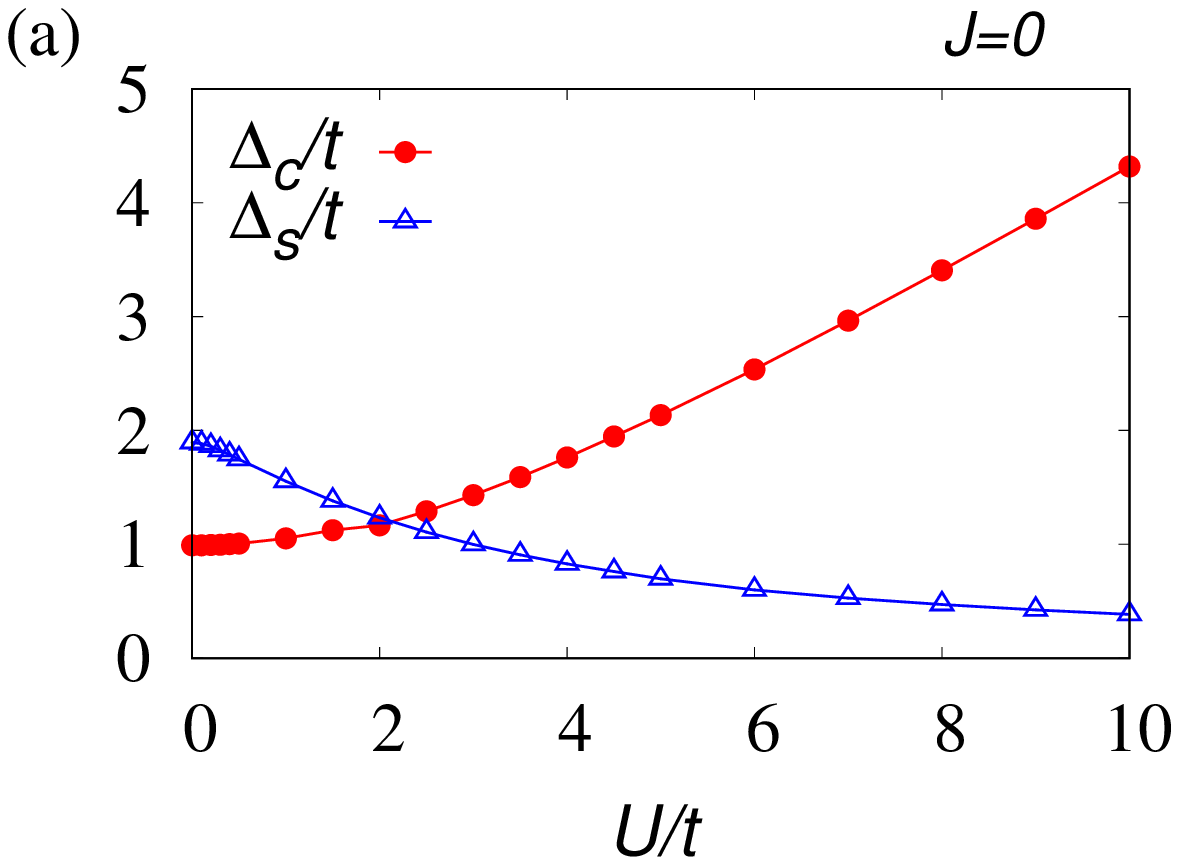}
\end{center}
\end{minipage}
\begin{minipage}{0.47\hsize}
\begin{center}
\includegraphics[width=50mm,clip]{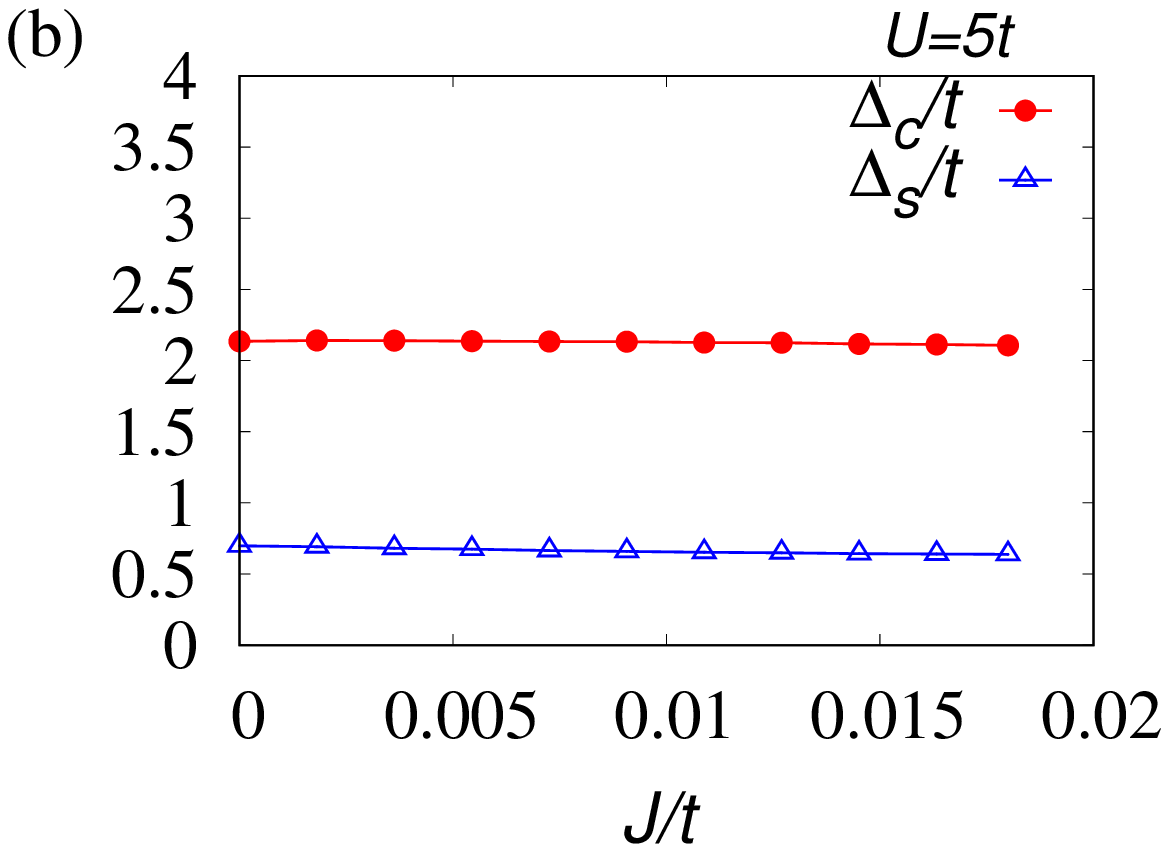}
\end{center}
\end{minipage}
\caption{(Color Online). 
The charge gap $\Delta_c$ and the spin gap $\Delta_s$ as functions of the interaction strength.
We have defined $\Delta_c$ and $\Delta_s$ as follows:
$\Delta_c=E_{2L+1,1/2}-E_{2L,0}$ and 
$\Delta_s=E_{2L,1}-E_{2L,0}$.
}
\label{fig: PBC gap}
\end{figure}

In order to discuss the bulk properties, we analyze the system under the PBC.
The charge gap and the spin gap are plotted for $0\leq U\leq 5t$ ($0\leq J\leq 0.018$ with $U=5t$) in Fig.~\ref{fig: PBC gap}(a) [Fig.~\ref{fig: PBC gap}(b)], respectively.

These data indicate that the bulk remains gapped along the path denoted in the phase diagram in Fig.~\ref{fig: OBC phase_ES}(b).

\section{Observation of another excitation gap}
\label{app: observation}
The argument around Eq.~(\ref{eq: At}) can be applicable to measuring the gap of triplet excitations conserving the $z$-component of the total spin, $E_{e,2L,0}-E_{2L,0}$,
where $E_{e,N,\tilde{S}^z}$ denotes the second lowest energy of the Hilbert space labeled by $(N,\tilde{S}^z)$.

To be concrete, the excitation gap can be measured as follows.
(i) A superposed state of the ground state and the triplet state with $\tilde{S}^z=0$ is obtained by applying a magnetic field gradient.
[Under the magnetic gradient $|\uparrow \rangle_a |\downarrow \rangle_b$ is preferred than $(|\uparrow \rangle_a |\downarrow \rangle_b - |\downarrow \rangle_a |\uparrow \rangle_b)/\sqrt{2}$, where $|\uparrow \rangle_{a(b)}$ denotes the state of the edge spin of channel $a$ ($b$), respectively.]
(ii)Oscillation is observed via $\langle \tilde{S}^z_a \rangle$. $\langle \tilde{S}^z_a\rangle$ can be obtained just by measuring $n_{a\uparrow}$ and $n_{a\downarrow}$.
We note $\langle singlet | \tilde{S}^z_a|triplet \rangle =1/2$.

\begin{figure}[!h]
\begin{minipage}{1\hsize}
\begin{center}
\includegraphics[width=50mm,clip]{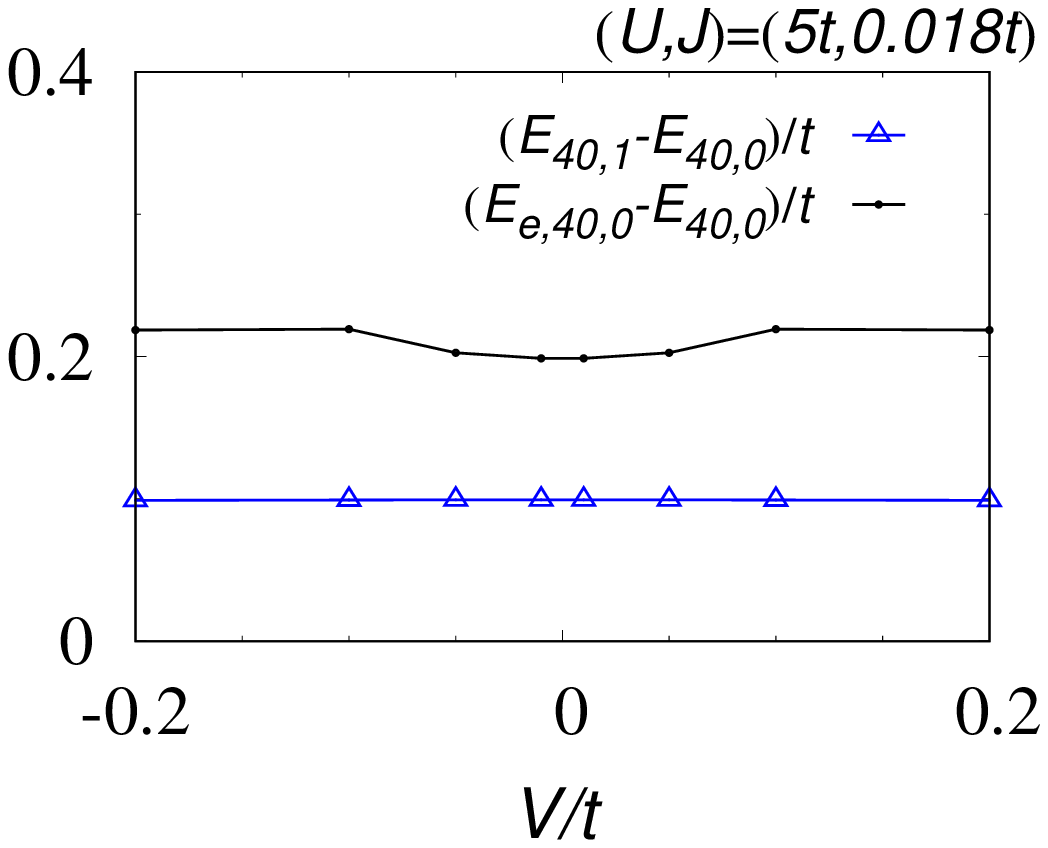}
\end{center}
\end{minipage}
\caption{(Color Online). 
$V$ dependence of the spin gap $\Delta_s:=E_{40,1}-E_{40,0}$ and the gap of the triplet excitation conserving the $z$-component of the total spin, $E_{e,2L,0}-E_{2L,0}$,
}
\label{fig: edge singlet-triplet}
\end{figure}

Our numerical data of the gap $E_{e,2L,0}-E_{2L,0}$ is summarized in Fig.~\ref{fig: edge singlet-triplet}.
From these data, we can estimate the size of the gap $E_{e,2L,0}-E_{2L,0}$ which is approximately $200\mathrm{Hz}$.
In Fig.~\ref{fig: edge singlet-triplet}, we can see that the gap $E_{e,2L,0}-E_{2L,0}$ is larger than the spin gap $\Delta_s$ for $-0.2t \leq V \leq 0.2t$, 
which indicates that the spin gap $\Delta_s$ is more relevant than the gap $E_{e,2L,0}-E_{2L,0}$ for the destruction of the gapless spin excitation.

\section{
Minimum energy of the Hilbert space of $(N,\tilde{S}_z)=(2L,1)$ and $(N,\tilde{S}_z)=(2L,-1)$
}
\label{app: Emin_up_down}
\begin{figure}[!h]
\begin{minipage}{0.45\hsize}
\begin{center}
\includegraphics[width=50mm,clip]{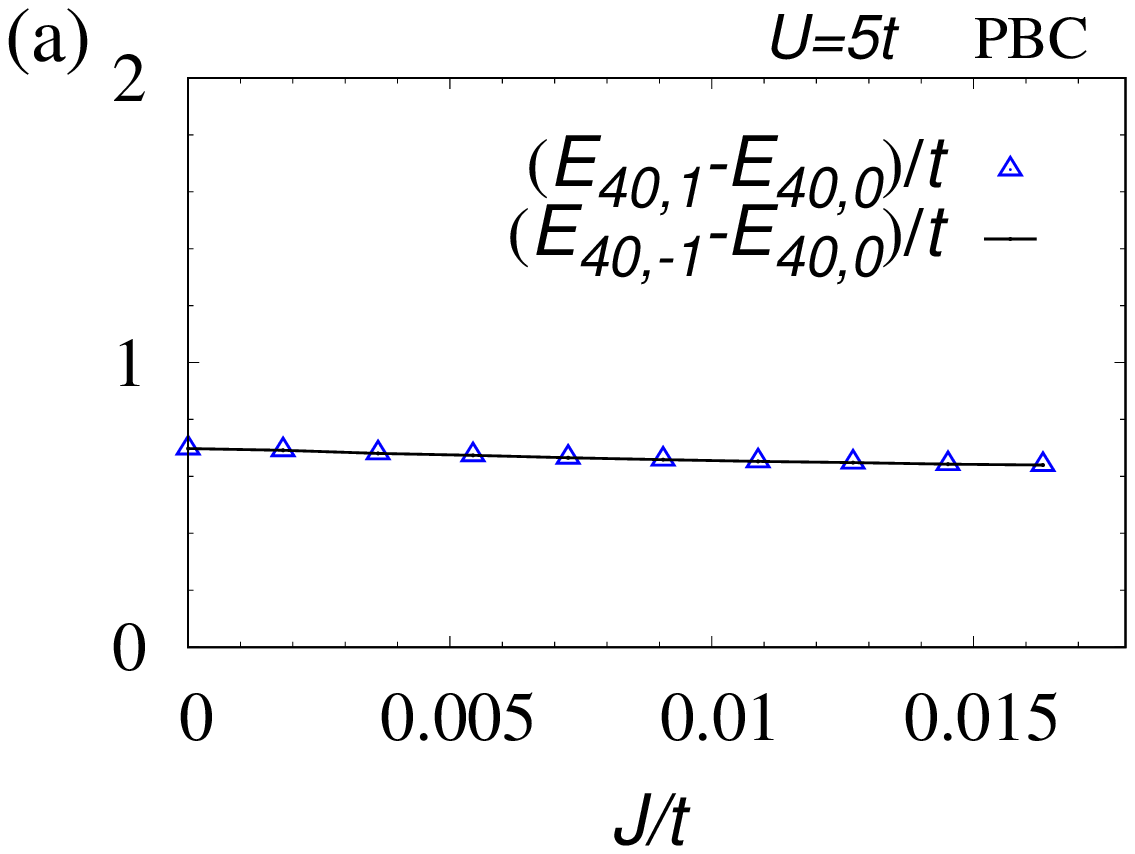}
\end{center}
\end{minipage}
\begin{minipage}{0.45\hsize}
\begin{center}
\includegraphics[width=50mm,clip]{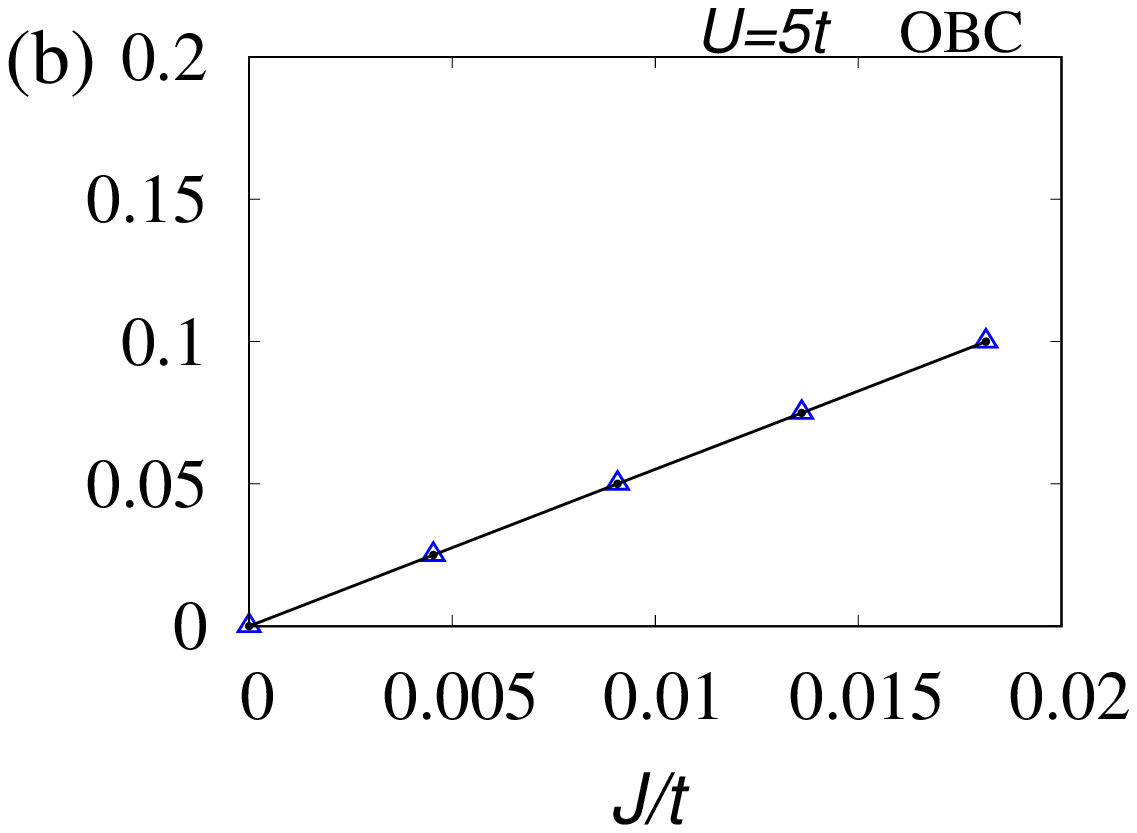}
\end{center}
\end{minipage}
\caption{(Color Online). 
Energy gaps for $U=5t$ as functions of $J$.
Data in panel (a) [(b)] are obtained under the PBC (OBC).
}
\label{fig: Szsymm gap}
\end{figure}

Here, we observe that for strong $U$, the gap of the up spin state becomes identical to that of the down spin state.
In Fig.~\ref{fig: Szsymm gap}(a), the excitation gaps of the up and the down spin states, $E_{2L,1}-E_{2L,0}$ and $E_{2L,-1}-E_{2L,0}$, are plotted under the PBC. Here, $E_{N,\tilde{S}^z}$ denotes the lowest energy in the Hilbert space labeled with $(N,\tilde{S}^z)$.
$N$ denotes the total number of fermions. $\tilde{S}^z$ denotes $z$-component of the total pseudo-spin. Fig.~\ref{fig: Szsymm gap}(a) indicates that both of the energy gaps take same values.

This is also the case under the OBC, which is shown in Fig. ~\ref{fig: Szsymm gap}(b).

\section{
Preparing the singlet state at edges.
}
\label{app: purifying}
The size of spin gap at edges is considered to be smaller than energy scale of temperature. Therefore, the singlet state at the edge is thermally mixed with excited states.

Here, we discuss how to prepare the purified singlet state at edges.
As shown in Fig.~\ref{fig: edge singlet-triplet}, the first excited states are triplet states with $\tilde{S}^z=\pm 1$.
The second excited states are the triplet state with $\tilde{S}^z=0$.
By making use of the feedback control, we can prepare the state $|\uparrow,\downarrow \rangle$ at the edge which is superposed state with the singlet state ($|0,0\rangle$) and the triplet state with $\tilde{S}^z=0$ ($|1,0\rangle$).
The corresponding wave function is written as
\begin{eqnarray}
|\psi(t=0)\rangle&=&|\uparrow,\downarrow\rangle, \\
&=& \frac{1}{\sqrt{2}}(|0,0\rangle +|1,0\rangle).
\end{eqnarray}
Such feedback control will be enabled if nondemolition measurement of local spin states with single-site resolution is realized~\cite{Nondemolition_Yamamoto_17}. 

After time-evolution with $t_0= \pi /[8 (E_t-E_s)] $, the state is written as
\begin{eqnarray}
|\psi(t_0)\rangle&=&
\frac{(e^{-iE_st_0}+e^{-iE_tt_0})}{\sqrt{2}}(
|\uparrow,\downarrow\rangle
-
i|\downarrow,\uparrow\rangle),
\end{eqnarray}
where $E_s$ and $E_t$ are energy of the singlet state and the triplet state.
Shining a pulse of magnetic field to this state, which adjusts the relative phase between  $|\uparrow,\downarrow\rangle$ and $|\uparrow,\downarrow\rangle$, yields the edge state 
$|\psi'(t_0)\rangle  \propto |singlet\rangle$.


\begin{thebibliography}{60}
\bibitem [{foo()}]{footnote_red}%
  \BibitemOpen
  \href@noop {} {} \bibinfo {note} {{ 
  Destruction of all the edge modes in this model indicates that the two-leg SSH model~(\ref{eq: free_SSH}) with the winding number four is topologically identical to the trivial phase
  }}\BibitemShut {NoStop}%
\bibitem [{foo()}]{footnote_pertub}%
  \BibitemOpen
  \href@noop {} {} \bibinfo {note} {{ The kinetic spin
  exchange interaction ($\sim t'{}^2/U$) requires local interchain hopping
  $t'$, which breaks the chiral symmetry }}\BibitemShut {NoStop}%
\bibitem [{foo()}]{footnote_SSH}%
  \BibitemOpen
  \href@noop {} {} \bibinfo {note} {{ The SSH model for
  optical lattice has been already created~\cite{Feolling_free_SSH07,
  Zak_phase_cold_atoms_Atala_NatPhys13}. The geometry of two-leg laddar can be
  created in experiments by means of the double-well optical
  lattice~\cite{Strabley_two-leg_06,Feolling_free_SSH07, Danshita_two_leg_07,
  Chen_two_leg_10} }}\BibitemShut {NoStop}%
\end{thebibliography}

\end{document}